\documentclass[a4paper,11pt]{article}

\usepackage{vmargin}
\usepackage{amsfonts}
\usepackage{amsmath}
\usepackage{amssymb}  
\usepackage{graphics,graphicx,color}
\usepackage{psfrag}

\newcommand{\beq}{\begin{equation}}
\newcommand{\eeq}{\end{equation}}
\newcommand{\beqa}{\begin{eqnarray}}
\newcommand{\eeqa}{\end{eqnarray}}
\newcommand{\nn}{\nonumber}
\newcommand{\R}{\mathbb{R}}

\newcommand{\cC}{\mathbb{C}}
\newcommand{\C}{\varphi}

\newcommand{\la}{\langle}
\newcommand{\ra}{\rangle}
\newcommand{\g}{\mathfrak{g}}
\newcommand{\h}{\mathfrak{h}}

\newcommand{\lalg}[1]{\mathfrak{#1}}  
\newcommand{\SU}{\mathrm{SU}}
\newcommand{\SO}{\mathrm{SO}}
\newcommand{\SL}{\mathrm{SL}}
\newcommand{\U}{\mathrm{U}}
\newcommand{\Spin}{\mathrm{Spin}}

\newcommand{\su}{\lalg{su}}
\renewcommand{\sl}{\lalg{sl}}

\newcommand{\so}{\lalg{so}}

\newcommand{\spin}{\lalg{spin}}

\newcommand{\Ad}{\mathrm{Ad}}
\newcommand{\End}{\mathrm{End}}

\DeclareMathOperator{\tr}{tr}

\setpapersize{USletter}
\setmarginsrb{39mm}{12mm}{28mm}{8mm}{12pt}{11mm}{0pt}{13mm}		 
    
\begin{document}

\sloppy
\title{\Large\bf Equivalence of the self-dual and Nambu-Goto strings}

\author{Winston J. Fairbairn \footnote{winston.fairbairn@uni-hamburg.de}\  
\\ [1mm]
\itshape{\normalsize{Department Mathematik, Universit\"at Hamburg,}} \\
\itshape{\normalsize{Bundesstra\ss e 55,  20146 Hamburg, Germany}} \\
\\ [1mm]
Karim Noui \footnote{Karim.Noui@lmpt.univ-tours.fr} , Francesco Sardelli \footnote{Francesco.Sardelli@lmpt.univ-tours.fr}\
\\ [1mm]
\itshape{\normalsize{Laboratoire de Math\'ematiques et Physique Th\'eorique}} \\
\itshape{\normalsize{Universit\'e Fran\c cois Rabelais, Parc de Grandmont, 37200 Tours, France}} }
\date{{\small\today}}
\maketitle

\abstract{We establish explicitely the relation between the algebraic and Nambu-Goto strings
when the target space is a four dimensional flat space. We find that the two theories are exactly equivalent only when the algebraic string is restricted to the self-dual or anti self-dual sectors. In its Hamiltonian formulation,
the algebraic string defines a constrained system with first and second class constraints. In the self-dual case,
we exhibit the appropriate set of second class constraints such that the resulting physical phase space is formulated 
in the same way as it is in the standard Nambu-Goto string. We conclude with a discussion on alternative quantisation
schemes.}

\section{Introduction}

The algebraic string is a first order formulation of the Nambu-Goto (NG) string introduced three decades ago by Balachandran, Lizzi, Sorkin and Sparano \cite{Bal2}. This formulation was later generalised to include spinning terms in \cite{Stern} and coupled to gravity in three and four space-time dimensions in \cite{Stern1,W}. The fields of the theory are maps from the string world-sheet to the isometry group of the maximally symmetric solution to Einstein's equations for the target space, that is, for zero cosmological constant, the Poincar\'e or Euclidean group depending on the signature of the target space metric. While the translational part consists in the standard embedding maps, the fields valued in the non-abelian part, i.e., the Lorentz or rotation group, are pure first order fields, the elimination of which reduces the algebraic string action to the NG string action. 

\medskip

This formulation of strings is interesting for several reasons. The action offers the advantage of being quadratic in the embedding maps and is thus better suited for path integral quantisation than the original NG action. Furthermore, it provides striking analogies with the first order formulation of four-dimensional gravity which makes it an ideal testing ground for ideas and techniques used in approaches attempting a quantisation of general relativity in four dimensions such as loop quantum gravity  \cite{Ash} and spin foam models \cite{SF}.

\medskip

The self-dual string \cite{Us} is a formulation of the algebraic string in terms of self-dual variables: the maps valued in the non-abelian part of the Poincar\'e or Euclidean group are restricted to lie in the self-dual or anti-self-dual subgroup. Remarkably, the self-dual variables are sufficient to describe the NG string in the same way as Ashtekar's variables \cite{selfdualgravity} are sufficient to capture the dynamics of full general relativity. 

\medskip

In \cite{Us}, a canonical analysis of the algebraic string was performed. The system was shown to admit first and second class constraints. The first class constraints were extracted and the Dirac bracket associated to the second class constraints was computed. It was shown that the self-dual action describes the same number of degrees of freedom as the NG string, namely two per worldsheet point. 
The description of the physical phase space, however, turned out to differ from the physical phase space of the NG string. In particular, the embedding maps were proven to be non-commutative in the Dirac bracket. In the non-self-dual case, the same analysis was carried out with a striking result; the general algebraic string describes one extra degree of freedom than the NG string.

\medskip

The canonical aspects of the algebraic string revealed in \cite{Us} thus lead to an apparent paradox. On the one hand, at the Lagrangian level, one can seamingly solve the the equations of motion for the second order fields appearing in the algebraic string action and evaluating the action on the solutions leads to the NG action. This shows that the algebraic and NG strings are classicaly equivalent. On the other hand, from the Hamiltonian perspective, there is a mismatch between the number of degrees of freedom described by the general algebraic string action and those described by the NG action. In the self-dual case, the numbers do match but the description of the physical phase space obtained in \cite{Us} differs from the standard physical phase space of the NG string.

\medskip

This paper is dedicated to shed some light on these apparent contradictions.  Besides a generalisation of the framework to both Euclidean and Lorentzian signatures for the target space metric, our main result is that the self-dual and Nambu-Goto strings are in fact exactly equivalent, while the general algebraic string theory differs generically from the NG string theory.

\medskip

In the first part of the article, 
we provide a careful analysis of the passage from first order to second order action in the Lagrangian framework. We find that there is in fact a degeneracy in the procedure; the algebraic string contains multiple sectors of solutions each corresponding to NG strings with different tensions. We find that, unless extra constraints are added by hand, as proposed in \cite{Bal2}, only the self-dual formulation is free of degeneracies and thus exactly equivalent to the NG string. This explains the anomaly in the number of degrees of freedom for the general algebraic string. In the second part of the article, we follow a different procedure to extract the first class constraints than in \cite{Us}. This new approach provides, via an appropriate choice of second class constraints, an exact equivalence between the obtained physical phase space and the phase space of the NG string. We also comment on the Hamiltonian aspects of the general algebraic string and give an interpretation for
the extra degree of freedom found in \cite{Us}. We conclude the article by discussing alternative quantisation procedures as a mean to test ideas and techniques used in four-dimensional quantum gravity. We propose a way to construct GNS states which are invariant under diffeomorphisms but weakly continuous at the same time.

\section{Lagrangian aspects of the algebraic string}

In this section, we present a generalised Lagrangian formulation of the algebraic string allowing us to unravel the different sectors contained by the theory. We argue that only the self-dual formulation is equivalent to the Nambu-Goto string  without additional constraints.

\subsection{First and second order forms of the algebraic string action}

Throughout the paper, $\spin(\eta)$ will denote the real Lie algebra of the isometry group $\Spin(\eta)$ associated to the flat metric $\eta = (\sigma^2,+,+,+)$, where $\sigma \in \{1,i \}$, that is the spin cover of the isometry group $\SO(\eta)$. The value $\sigma = i$ (resp. $\sigma = 1$) therefore corresponds to Lorentzian (resp. Euclidean) signatures in which case $\spin(\eta) =\spin(3,1)$ (resp. $\spin(\eta) = \spin(4)$). The vector representation of $\Spin(\eta)$ will be noted $(\pi,V_{\eta})$ and we will chose a basis $(e_I)_{I=0,...3}$ of $V_{\eta}$.

We will consider a (non-critical) closed spinless bosonic string propagating on a four-dimensional flat manifold $\mathbb M \cong V_{\eta}$. Let $\Sigma \subset \mathbb M$ be the string world-sheet, i.e., a two-dimensional closed, compact, oriented sub-manifold of $\mathbb M$. The string is described by a pair of fields $\Lambda=(X,g)$ on the string world sheet $\Sigma$; $X$ is the embedding map $X : \Sigma \rightarrow \mathbb M$, 
and the field $g$ is a smooth map $g : \Sigma \rightarrow \Spin(\eta)$ valued in the spin group.

The basic building block for the algebraic string action is the two-form $dX \wedge dX$. Introducing local coordinates $(dx^{\alpha})_{\alpha=0,1}$ 
on the co-tangent bundle $T^* \Sigma$ and the two-dimensional Levi-Cevita tensor $\epsilon^{\alpha \beta}$ normalised by $\epsilon_{01} = 1$, the two-form $dX \wedge dX$ is given locally by
\beq
d X^{I} \wedge dX^{J} = B^{IJ} d^2 x, \;\;\;\;\; B^{IJ} := \epsilon^{\alpha \beta} \partial_{\alpha} X^{I} \partial_{\beta} X^{J},
\eeq
where the map  $B : \Sigma \rightarrow \Lambda^2(V_{\eta})$ is called the $B$ field. We will assume that the condition $\sigma^2 \langle B , B \rangle > 0$ ($B$ is a time-like 
bivector in Lorentzian signatures) is satisified.
The bilinear form $\langle , \rangle$ on $\spin(\eta)$ is the Killing form (see the Appendix for a detailed account of our conventions).

The algebraic string action \cite{Bal2} depends on the variables $X$ and $g$ and is given by 
\beq
\label{general}
S[X,g] = \int_{\Sigma} \la M, dX \wedge dX \ra =  \int_{\Sigma} d^2 x \, M_{IJ} \, B^{IJ}.
\eeq
Here, $M=\Ad_{g^{-1}}(k) = g^{-1} k g$, with $k$ a fixed element in $\spin(\eta)$ which parametrises the set of theories described by the action.

The above action admits the following symmetries. It is globally Poincar\'e (or Euclidean) invariant. It also admits two types of gauge symmetries. It is invariant under the isotropy group of $k$ in $\Spin(\eta)$ and under the diffeomorphism group Diff($\Sigma$) of the worldsheet.

The free parameter $k$, or more precisely the conjugacy class to which $k$ belongs, can be tuned such that the above action contains the Nambu-Goto string. To understand this last point, we perform the variation of the action with respect to the group variable $g$. Using the right variation 
$$
\delta g = g \circ T, 
$$
with $T$ an arbitrary element of $\spin(\eta)$, together with the invariance of the Killing form, we obtain the equations of motion
\beq
\label{motiong}
\delta S = 0 \;\;\;\; \Leftrightarrow \;\;\;\; [M,B] = 0.
\eeq
Thus, the motion forces $M$ to lie in the centraliser $C(B)$ of $B$ in $\spin(\eta)$. 
Because of the rank of $\spin(\eta)$, $C(B)$ is generically a two-dimensional (real) vector space spanned
by $B$ and $\star B$. Therefore, the action is extremal if and only if
\beq
\label{M}
M = \alpha B + \beta \star B, \;\;\;\; \alpha,\beta: \Sigma \rightarrow \R,
\eeq
where $\star : \spin(\eta) \rightarrow \spin(\eta)$ is the Hodge linear map. The solutions to the equations of motion are therefore characterised by the values of $\alpha$ and $\beta$. These parameters are in turn determined, up to degeneracies, by the conjugacy class of $k$, or equivalently of $M$. The conjugacy class of $M$ in $\spin(\eta)$ is totally caracterised by the two adjoint action invariants constructed from the two Ad-invariant bilinear forms on $\spin(\eta)$ (see the Appendix)
\beq
\tau(M) = \la M^2 \ra \;\;\;\; \mbox{and} \;\;\;\; s(M) = ( M^2 ),
\eeq
where we have introduced the notation $\la X^2 \ra = \la X , X \ra$ and $( X^2 ) = ( X , X )$ for all $X$ in $\spin(\eta)$. Note that the above invariants satisfy $\tau(\pm \star M) = \sigma^2 \tau(M)$ and $s(\pm \star M) = \sigma^2 s(M)$ which is the root of the degeneracies discussed below. From here on, for notational convenience, $\tau(M)$ and $s(M)$ will simply be noted $\tau$ and $s$. 

Using the fact that $( B^2 ) =0$, i.e. the bivector $B$ is {\em simple}, one can calculate the two invariants $\tau$ and $s$ as functions of $B$
\beq
\tau = (\alpha^2 + \sigma^2 \beta^2) \langle B^2 \rangle, \;\;\;\;\;\;\;\; s = 2 \sigma^2 \alpha \beta \langle B^2 \rangle.
\eeq
One can next invert this system to obtain the two unknowns $\alpha,\beta$ as functions of the invariants $\tau$ and $s$, that is, as functions of the conjugacy class of $k$. Given $\tau$ and $s$, there are generically four different pairs of variables $\alpha$ and $\beta$ satisfying the above equation because the relations are quadratic in $\alpha$ and $\beta$. Introducing a pair of signs $\varepsilon, \varepsilon' = \pm 1$, the four possible solutions for $\alpha$ and $\beta$ respectively denoted $\alpha_{\epsilon, \epsilon'}$ and $\beta_{\epsilon, \epsilon'}$, are given by
\beq
\label{alpha}
\alpha_{\epsilon, \epsilon'} = \frac{\varepsilon a_{+} + \varepsilon' a_{-} }{2 \sigma \sqrt{\sigma^2 \la B^2 \ra}}, \;\;\;\;\;
\beta_{\epsilon, \epsilon'} = \frac{\varepsilon a_{+} - \varepsilon' a_{-} }{2 \sqrt{ \sigma^2 \la B^2 \ra}} ,
\eeq
where $a_\pm = \sqrt{\tau \pm \sigma s}$ are defined with a given choice of a square root.

The space of solutions to the equations of motion \eqref{motiong} for $g$ therefore consist of four sectors. Labeling the four possible sectors by  $(\epsilon, \epsilon')$, a given solution is necessarily of the form
$$
M_{\epsilon, \epsilon'} = \alpha_{\epsilon, \epsilon'} B + \beta_{\epsilon, \epsilon'} \star B.
$$
There is a $\mathbb{Z}_2 \times \mathbb{Z}_2$ symmetry $M \mapsto \star M$, $M \mapsto -M$ on the space of solutions which maps a solution in one sector to a solution in another sector as follows
\beq
\label{symmetry}
\star M_{\epsilon, \epsilon'} = \frac{1}{\sigma} M_{\epsilon, -\epsilon'}, \;\;\;\; - M_{\epsilon, \epsilon'} =  M_{-\epsilon, -\epsilon'}.
\eeq
This is reminiscent to the (non-self-dual) Plebanski formulation of four-dimensional gravity \cite{Pleb} as remarked in \cite{FdP}.

Using the simplicity of the $B$ field, it is immediate to see that the evaluation of the algebraic string action \eqref{general} on any of the $(\epsilon, \epsilon')$ sectors does not depend on the $\beta$ variable and the obtained second order action yields
\beq
\label{2ndorder}
S_{\epsilon, \epsilon'}[X] = C_{\epsilon, \epsilon'}(k) \, \int_{\Sigma} d^2 x \, \sqrt{ \sigma^2 \la B^2 \ra}, \;\;\; \text{with} \;\;\;
C_{\epsilon, \epsilon'}(k)=\frac{\sigma}{2}(\varepsilon a_+ + \varepsilon' a_-) \,.
\eeq
The relation between the B field and the induced metric on the worldsheet reads
$$
 \la B^2 \ra = \det X^* \eta,
$$
where the star $*$ is the pull back map,
and the algebraic string evaluated on any of the $(\epsilon, \epsilon')$ sectors yields the Nambu-Goto action 
\beq
S_{\epsilon, \epsilon'}[X] = C_{\epsilon, \epsilon'}(k) \, \int_{\Sigma} d^2 x \, \sqrt{\sigma^2 \det X^* \eta},
\eeq
provided the coupling constant $C_{\epsilon, \epsilon'}(k)$, that is, the conjugacy class of $k$, is chosen to equate the tension of the string, that is, chosen such that $C_{\epsilon, \epsilon'}(k)2\pi\alpha'=1$, with $\alpha'$ the Regge slope. Therefore, the $(\epsilon, \epsilon')$ sector of the algebraic string is classically equivalent to the NG string with Regge slope $\alpha'$ provided the above condition is satisfied. However, the full theory contains four such sectors with different coupling constants and thus generically the algebraic string is not equivalent to the NG string, unless some extra constraints are imposed as remarked in the original paper \cite{Bal2}.  Therefore, it turns out that the conjugacy class of $k$ does not parametrise the set of theories appropriately; a particular choice of a conjugacy class of $k$ leads to four possible string theories.

\subsection{Study of the degeneracies: emergence of the self-dual string}

To have a satisfactory classical equivalence with the NG string, one would need a prescription to uniquely chose and isolate only one sector of solutions. To this aim, we study the equations satisfied by the coupling constant $C(k)$, or equivalently by $\alpha$, and show that there exists a natural framework in which the four sectors of solutions collapse to two sectors related by a sign. In other word, the symmetry on the space of solutions \eqref{symmetry} reduces to a single $\mathbb{Z}_2$ symmetry which will turn out to play no role in the dynamics of the theory. This framework is that of the self-dual string discovered in \cite{Us}.

 Let $(\epsilon, \epsilon')$ be a given sector of solutions. Studying the structure of $C_{\epsilon, \epsilon'}$ displayed in \eqref{2ndorder} leads to the conclusion that the (squared) coupling constant satisfies the following polynomial equation
\beq
\label{equation for C}
\sigma^2 \, C_{\epsilon, \epsilon'}(k)^4 \, - \, \tau \, C_{\epsilon, \epsilon'}(k)^2 \, + \, \frac{s^2}{4} \, = \, 0 
\eeq
which admits an unique  solution for $C_{\epsilon, \epsilon'}(k)^2$ if and only if
\beq\label{unicity}
\tau \,  + \, \sigma \, s \, = \, 0 \;\;\;\; \mbox{or} \;\;\;\; \tau \,  - \, \sigma \, s \, = \, 0.
\eeq
In this case, the four sectors of solutions collapse to two sectors and the two possible coupling constants are given by $C_{\epsilon, \epsilon'}(k) = \pm \sqrt{\tau/2 \sigma^2}$.

The unicity conditions \eqref{unicity} can be given a nice algebraic interpretation. Let $M=M_++M_-$ be the decomposition of $M$ into self-dual and anti self-dual components $M_\pm$ (see Appendix). It is immediate to calculate the invariants $\tau$ and $s$ in terms of $M_{\pm}$ from which we obtain
\beq
\tau \, \pm \, \sigma \, s \; = \; (1\pm \sigma^2) \la M_+^2 \ra \; + \; (1\mp \sigma^2) \la M_-^2 \ra \,.
\eeq

Therefore, the condition (\ref{unicity}) holds if and only if $\la M_+^2 \ra=0$ or $\la M_-^2 \ra=0$ which is equivalent to saying that
$M$ is self-dual or anti self-dual. In this case, the algebraic string action reduces to its self-dual or anti-self-dual formulation. In the self-dual case, the action reads
\beq
\label{selfdual}
S[X,g_+] = \frac{1}{2} \int_{\Sigma} d^2 x \, M_{+ \, IJ} \, B^{IJ}_+,
\eeq
where $g_+$ belongs to the self-dual subgroup $H_+(\sigma)$, where $H_+(1) \cong \SU(2)$ and $H_+(i) \cong \SL(2,\cC)$ (see the Appendix for a precise account of our conventions). If we chose the conjugacy class of $k$ such that $\tau = \sigma^2 /2(\pi \alpha')^2$, e.g.  $k = (\sigma / \pi \alpha') T_{+3}$ with $(T_{+a})_{a=1,2,3}$ the basis of the self-dual algebra $\mathfrak{h}_+(\sigma)$ chosen in the Appendix, the second order form of the action \eqref{2ndorder} reads exactly the NG action with the degeneracy reduced to a sign.

Note that $\tau <0$ in Lorentzian signatures. As a consequence, $M$ is necesseraly a complex
variable in the Lorentzian regime and a priori reality conditions have to be imposed to recover the real theory.
However, the Hamiltonian analysis will show that the $M$ variables can be solved  in terms of the $X$ variables and thus disappear from the formalism.

We conclude by discussing the original proposal \cite{Bal2}. In the original paper, the class of $k$ is chosen such that $\tau = \sigma^2 / (2 \pi \alpha')^2$ and $s=0$, e.g. $k = 1/(2 \pi \alpha') T_{03}$ with $(T_{\alpha \beta})_{\alpha<\beta = 0, ... , 3}$ the set of generators of $\spin(\eta)$ chosen in the Appendix. From the polynomial equation above \eqref{equation for C}, we read that this choice yields two possible coupling constants up to a sign; the first is the string tension as desired while the second solution is zero. The degeneracy is dealt with by adding by hand a strict positivity constraint to the Lagrangian ($\sigma^2 \mathrm{Tr} M B > 0$) as a mean to suppress all unwanted solutions. This explains why the canonical analysis performed in \cite{Us} revealed an anomaly in the number of degrees of freedom in the theory defined by the general algebraic string action; the positivity constraint was not implemented in the Hamiltonian framework which means that the
  analysis described a theory in a sense larger than the NG string theory. We will come back to this point in the sequel. 

\section{Hamiltonian aspects of the algebraic string}

In this section, we perform a full canonical analysis of the self-dual string and provide a proof of the equivalence to the Hamiltonian NG string. This is achieved by following a different strategy than the one followed in \cite{Us} to extract  the first class constraints of the system. This leads to another natural choice of second class constraints which allows us to show the equivalence between the physical phase spaces of the two theories. We then comment on the general algebraic string and make contact with the Lagrangian aspects.

\subsection{Hamiltonian analysis of the self-dual string}

The Hamiltonian setting is as follows.
We suppose that the world sheet $\Sigma$ is diffeomorphic to the cylinder and foliate it by a one parameter family of one-dimensional `spatial' manifolds $S_t$, $t \in \R$, each diffeomorphic to the circle, that is, $\Sigma \simeq \R \times S$. Let $x \in [0,2 \pi]$ denote a parametrisation of the circle $S$ and let the configuration variables satisfy $X(t,0) = X(t,2\pi)$ and $g(t,0) = g(t,2\pi)$ for all $t$ in $\R$. 

The canonical decomposition of the algebraic string action \eqref{general} yields
\beq
\label{canonical}
S = 2 \int_{\R} dt \, \int_{S} dx \, M_{IJ} \, \dot{X}^{I} \, X'^{J},
\eeq
where we have introduced the standard notation $\dot{\phi} \equiv \partial_t \phi$, and $\phi' \equiv \partial_x \phi$ for all field $\phi$ on the world-sheet.
The self-dual action has exactly  the same form but $M$ is self-dual, i.e., $M=M_+$. In the following, we will omit for simplicity the index $+$ to 
specify the self-dual components. We will adopt this simplification in the whole section as we are considering only the self-dual
action here and there is no possible confusion.
From this canonical action, we can read out the momenta conjugate to the configuration variables $(X,g)$ and study the constraints of the system. 

\subsubsection{The set of constraints}

We start by introducing the momenta $\pi_I$, $I=0,...,3$ conjugate to the variables $X^I$. The corresponding symplectic structure is read out of the Poisson brackets
\beq\label{bracketPiX}
\{\pi_I(x), X^J(y) \} = \delta_I^J \, \delta(x,y).
\eeq
For the group-valued variables, we will consider the parametrisation of the cotangent bundle $T^*H_+(\sigma)$ in the left trivialisation (see e.g. \cite{Symp}) in terms of the canonical pair $(g,(P_a)_{a=1,2,3})$. The corresponding canonical symplectic structure leads to a Poisson structure, the only non-vanishing brackets being given by
\beq
\{P_a(x), g^I_{\;J}(y) \} = -(T_a \, g)^I_{\;J}(x) \, \delta(x,y), \;\;\;\; \{P_{a}(x), P_{b}(y) \} = \epsilon_{ab}^{\;\;\,c} \; P_c\, \delta(x,y),
\eeq
where $T_a := T_{+a}$ generate the self-dual Lie algebra $\h_+(\sigma)$. It is clear from the first Poisson bracket that $P_a$ are interpreted 
as left-derivatives on $H_+(\sigma)$.  

\medskip

Together with (\ref{bracketPiX}), these Poisson brackets define the non-physical phase space of
the theory. It is non-physical because the canonical action \eqref{canonical} defines a constrained system since the conjugate momenta are fixed by 
the following set of seven primary constraints
\beqa
\label{constraints}
\varphi_I &:=& \pi_I - M_{IJ} \, X'^J \approx 0 \nn \\
\phi_a &:=& P_{a} \approx 0.
\eeqa
One can show that there are no further (secondary or tertiary) constraints in the system (see \cite{Us}).
The Poisson brackets between the constraints are given by
\beqa\label{systemofsecondclass}
\{ \C_I(x), \C_J(y) \} &=& - M'_{IJ}(x) \delta(x,y) \nn \\
\{ \C_I(x), \phi_a(y) \} &=& (g^{-1}[T_{a},k]g)_{IJ} X'(x)^J \delta(x,y) \nn \\
\{ \phi_a(x), \phi_b(y) \} &=& \epsilon_{ab}^{\;\;\,c} \; \phi_c(x) \, \delta(x,y).
\eeqa
The above calculation implies that the constraint matrix
\beq
\label{constmat}
C :=  \left( \begin{array}{cc} 
\{ \C_I(x), \C_J(y) \} & \{ \C_I(x), \phi_b(y) \} \\ 
\{ \phi_a(x), \C_J(y) \} & \{ \phi_a(x), \phi_b(y) \}
      \end{array} \right),
\eeq
is not weakly vanishing. Therefore, $(\varphi_I,\phi_a)_{I,a}$ does not form a set of first class constraints. 
However, these constraints are not {\em pure} second class, that is, the constraint matrix has a non-trivial kernel and is therefore not invertible. 
This comes from the fact that there are gauge symmetries in the system and thus this matrix has to admit non-trivial null vectors which are the first class 
constraints generating the symmetries.

\medskip

There are two canonical ways of extracting the first class constraints of the system. The first, that we followed in \cite{Us}, consists in first isolating a subset of pure second class constraints out of the whole set of constraints and then computing the associated (partial) Dirac bracket. The corresponding constraints can then be solved explicitly and one is left with a smaller set of constraints and the (partial) Dirac bracket as Poisson structure. One then repeats the procedure recursively until all second class constraints are exhausted. The remaining constraints are therefore first class in the Dirac bracket corresponding to the last step of the recursion. 

\medskip

The second method consists in explicitly computing the kernel of the constraint matrix which amounts to directly exhibiting the first class constraints. One can then find an equivalent description of the constraint surface where first and second class constraints are clearly separated. Although this procedure can always be implemented in principle, it can prove difficult in practice. In our case, the method can be carried out explicitly and offers the advantage of leading to a description of the physical phase space of the self-dual string in which the relation to the Nambu-Goto
physical phase space is crystal clear. We now implement the second method and compute the kernel of the constraint matrix.

\subsubsection{Extracting the first class constraints}

Throughout this section, quantities of the form $\eta^{IJ} f_J \C_I$ and $\delta^{ab} g_b \phi_a$ with $f$ valued in $\mathbb{M}$ and $g$ valued in  $\mathfrak{h}_+(\sigma)$ will often occur, in which case the shorthand notation $\C(f)$ and $\phi(g)$ will be employed.

To extract the first class constraints out of the above system, we compute the kernel of the constraints matrix, that is, determine linear combinations of the form
$$
\psi= \mu^I \C_I + \nu^a \phi_a \; := \; \C(\mu) + \phi(\nu) \, , 
$$
where  $\mu, \nu$ are phase space functions which are such that 
$$
\{ \psi, \C_I \} \approx 0 \;\;\;\; \mbox{and} \;\;\;\;  \nn \\
\{ \psi, \phi_a \} \approx 0.
$$

The study of the structure of the system of equations \eqref{systemofsecondclass} leads immediately to the identification of a first solution: it is simply given by taking $\mu^I=0$ and $\nu^a = k^a$, and the constraint 
\beq
H \, := \, \phi(k) =  k^a \phi_a
\eeq
 is in the kernel of the constraint matrix i.e., is first class. 

To determine the other solutions, a little more work is required. As a first step, using 
\beq
M' =  (g^{-1})' k g +  g^{-1} k g' = g^{-1} [k, g' g^{-1}] g,
\eeq
we remark the following identity
$$
\{ \C_I, \C_J \} X'^J = \{ \C_I, \phi_a \} ( g' g^{-1})^a.
$$
This implies immediately that the constraint 
\beq
H_1 := \C(X') - \phi(g' g^{-1}) = X'^I \C_I - ( g' g^{-1})^a \phi_a
\eeq
 is first class, as can be seen from the following calculations
$$
\{ H_1, \C_I \} \approx  \{ \C_J, \C_I \} X'^J -  \{\phi_a, \C_I \} ( g' g^{-1})^a = 0,
$$
and
$$
\{ H_1, \phi_a\} \approx (g^{-1}[T_{a} , k]g)_{IJ} \, X'^I X'^J = 0.
$$

\medskip

To find the next first class constraint, we firstly work out the following weak equality
\beq
\{ \C_I, \C_J \} \pi^J \approx - M_{IJ}' M^J_{\;\; K} X'^K = (g^{-1} [g' g^{-1},k] k g)_{IK} X'^K.
\eeq
Using equation \eqref{identity} of the Appendix, it is easy to derive that
\beq
[\lambda,k] k 
= \frac{1}{2} [[\lambda,k] k],
\eeq
for all matrix $\lambda$ in the vector representation of the self-dual subalgebra. It follows immediately that
\beq
\{ \C_I, \C_J \} \pi^J \approx \frac{1}{2} \{ \C_I, \phi_a \} [ g' g^{-1}, k]^a.
\eeq

As a consequence, it is immediate to see that 
\beq
\tilde{H}_0 := 2 \C(\pi) - \phi([ g' g^{-1}, k]) = 2 \pi^I \C_I - [ g' g^{-1}, k]^a \phi_a,
\eeq
is a first class constraint because
$$
\{ \tilde{H}_0, \C_I \} \approx 2 \{ \C_J, \C_I \} \pi^J -  \{\phi_a, \C_I \} [ g' g^{-1}, k]^a = 0,
$$
and
$$
\{ \tilde{H}_0, \phi_a\} \approx - (g^{-1}[[T_{a} , k],k]g)_{IJ} \, X'^I X'^J = 0.
$$

\medskip

Fo reasons that will become clear in the sequel, we will substract the quadratic combination $\C^I \C_I$ to $\tilde{H}_0$ and consider the first class constraint
\beq
H_0 = (2 \pi^I - \C^I)\C_I - [ g' g^{-1}, k]^a \phi_a.
\eeq

We have now identified three first class constraints. From the Lagrangian analysis, we know that there are three gauge symmetries in the system, the $\U(1)$ transformations stabilising $k$, and the two worldsheet diffeomorphims. Therefore, assuming that there are no accidental symmetries, we have extracted all the first class constraints of the self-dual algebraic string.

We conclude this section with a remark. Note that the constraints $H_1$ and $H_0$ can be rewritten as
\begin{eqnarray}
\label{H0H1}
H_1 & = & \pi_I X'^I - \phi(g' g^{-1}) \\
H_0 & = & \left(\pi_I \pi^I - \frac{\sigma^2}{(2 \pi \alpha')^2} X'_I X'^I \right) - \phi([ g' g^{-1}, k])
\end{eqnarray}
where we have used that  $M^2 = - \sigma^2 / (2\pi \alpha')^2 1\!\! 1$. We therefore recognise the standard diffeomorphism constraints of the Nambu-Goto and Polyakov strings (see e.g. \cite{books} and references therein), augmented by specific combinations of the $\phi_a$ constraints.

\medskip

Before computing the Dirac bracket for the second class constraints, we will first compute the action of the first class constraints on the phase space variables, and the Dirac algebra of the first class constraints. These calculations will not be affected by the introduction of the Dirac bracket because the Dirac bracket between a first class constraint and an arbitrary phase space function reduces to the original Poisson bracket.

\subsubsection{Symmetries: status and properties}

In this section, we explicitly compute the phase space flow of the first class constraints and check that these constraints generate the infinitesimal gauge symmetries of the system. To prove this point, we introduce the smeared version of these constraints  
\beq
\label{firstclassconstraints}
H(\alpha) \; = \; \int_S dx \,\alpha(x) \,H(x), \;\;\;\; H_1(u) = \int_S dx \, u(x) H_1(x), \;\;\;\;
 H_0(v)=\int_S dx \,v(x) H_0(x),
\eeq
with $\alpha, u$ and $v$ in $C^{1}(S,\R)$ are arbitrary functions independent of the dynamical variables of the theory. Let $f$ be a $C^1$ phase space function. We will make use of the notation $\delta_{w} f =  \{ C(w) , f \}$, with $w= \alpha, u,$ or $v$ and $C=H,H_1$ or $H_0$.

One can show that the action of $H$ yields
\beq
\delta_{\alpha} X^I = 0, \;\;\;\; \;\;\;\; \delta_{\alpha} \pi_I = 0, \;\;\;\; \;\;\;\;  \delta_{\alpha} g =  -\alpha kg, 
\eeq
while $H_1$ induces the transformations
\beq
\delta_{u} X^I = u X'^I, \;\;\;\; \;\;\;\; \delta_{u} \pi_I = (u \pi_I)', \;\;\;\; \;\;\;\; \delta_{u} g = u g'.
\eeq
Finally, $H_0$ generates the following action
\beq
\delta_{v} X^I = 2 v \pi^I, \;\;\;\; \;\;\;\; \delta_{v} \pi_I = - \frac{2 \sigma^2}{(2 \pi \alpha')^2} (v X'_I)', \;\;\;\; \;\;\;\; \delta_{v} g = - v g M' \,.
\eeq
From the above, we can immediately conclude that $H(\alpha)$ generates infinitesimal left $\U(1)$ transformations stabilising $k$ with parameter $\alpha$, and that $H_1(u)$ generates infinitesimal diffeomorphisms of the circle with vector field $u \, \partial_x$. By elimination, the last first class constraint $H_0(v)$ out of the three necessarily generates infinitesimal diffeomorphisms in the time-like direction, i.e., time reparametrisation. 

\medskip

To confirm this last point, we can compute the Hamilton equations of motion associated to the Hamiltonian $H_0$ for $X$ and $\pi$
\beqa
\label{motion}
\dot{X}^I &:=& \{H_0 , X^I \} = 2 v \pi^I, \nn \\
\dot{\pi}_I&:= &\{H_0 , \pi_I \} = - \frac{2 \sigma^2}{(2 \pi \alpha')^2} (v X'_I)' \nn
\eeqa
From the above equations, we can calculate the second time derivative of the $X$ variable 
\beq
\ddot{X}^I = - \frac{4v \sigma^2}{(2 \pi \alpha')^2} \left( v' X'_I + v X''_I \right) \nn
\eeq
Accordingly, we obtain the standard wave equation for $X^I$
\beq
\left(\sigma^2 \partial_t^2 + \partial_x^2 \right) \, X^I = 0,
\eeq
upon the choice of gauge $v = \pi \alpha'$.
This result ensures that $H_0$ can be interpreted as the Hamiltonian constraint of the theory.

\medskip

Is is clear that any definite choice of arbitrary multipliers in the total Hamiltonian is equivalent to a gauge condition, as it represents a restriction on the form of the general reparametrisations. In order to find the geometric meaning of the condition $v = \pi \alpha'$, we use the equation of motion of the embedding variables and implement the above choice of gauge on the constraint hypersurface determined by the $\C_I$
$$
\dot{X}^I  \approx 2 \pi \alpha' M_{\; J}^I X'^J.
$$
From this equation, we can calculate the components of the induced metric
\beq 
h_{\alpha \beta} = \partial_{\alpha} X^I \partial_{\beta} X_I .
\eeq
The calculation proceeds as follows. The diagonal components are related because
\beq
h_{00} \approx (2 \pi \alpha')^2 M_{\; J}^I X'^J M_{IK} X'^K = \sigma^2 h_{11} \nn
\eeq
where we have inserted the value of $M^2$ in the last equality. The off diagonal terms vanish as shown below
\beq
h_{01} \approx 2 \pi \alpha' M_{\; J}^I X'^J X'_I = 0. \nn
\eeq
The above results imply that, in the gauge defined by $v = \pi \alpha'$, the worldsheet metric is conformally flat
\beq
ds^2 = h_{\alpha \beta} \, dx^{\alpha} dx^{\beta} = e^{\phi} (\sigma^2 dt^2 + dx^2),
\eeq
where $\phi = \ln h_{11}$ is the conformal factor.

\medskip

We now display the Dirac algebra satisfied by the (smeared) constraints  \eqref{firstclassconstraints}. A straight-forward but lengthy calculation shows that
\beqa
&&\{H(\alpha),H(\alpha')\} \, = \, 0 \, , \,\, 
\{H(\alpha),H_1(u)\} \, = \, -H(u \alpha') \, , \,\, 
\{H(\alpha),H_0(v)\} \, = \, 0 \nn \\ 
&&\{H_1(u_1),H_1(u_2)\} \, = \, 2 H_1(u_1u_2'-u_2u_1')\label{H1H1} \, , \,\,
\{H_1(u),H_0(v)\} \, = \, 2 H_0(uv' - vu')\label{H1H0} \nn \\
&&\{H_0(v_1),H_0(v_2)\} \, = \,-\frac{2 \sigma^2}{(2\pi\alpha')^2} H_1(v_1v_2'-v_2v_1') + \tilde{H}(v_1v_2'-v_2v_1')\label{H0H0}, \nn 
\eeqa 
where $\tilde{H}(x)=H(x) \tr(k g' g^{-1})$ and $\tilde{H}(v)=\int_S dx\,v(x)\tilde{H}(x)$ the associated smeared function. The above computation is performed by remarking that all constraints are of the form $C_1(X,\pi) + C_2(g,P)$. 
Since $X$ and $\pi$ both commute with $g$ and $P$, the computation of the brackets splits into two separate, commuting parts. The $C_1$ part produces the standard results for the constraints algebra of the bosonic string, while the $C_2$ part was calculated in \cite{Us}. 

\medskip

We conclude this section discussing some aspects related to the observables of the theory.
The global Poincar\'e (or Euclidean) symmetry of the algebraic string action is reflected in the conservation of a momentum and angular momentum charge. It is immediate to show that the quantities
\beq
\Pi_I = \int_S dx \, \pi_I, \;\;\;\; \mbox{and} \;\;\;\; J^{IJ} = \int_S dx \left[ X^I \pi^J - \pi^I X^J \right],
\eeq
have vanishing Poisson brackets with all the constraints and are accordingly strong Dirac observables. Furthermore, they satisfy the Poisson algebra of the Poincar\'e (or Euclidean) group.

\subsubsection{Dirac bracket and physical phase space}

We have extracted the first class constraints from the initial set of constraints $(\C_I,\phi_a)_{I,a}$ by exhibiting linear combinations of the constraints which were shown to lie (weakly) in the center of the algebra of the constraints. We have therefore found an equivalent description of the constraint surface in terms of first class constraints $H,H_1,H_0$ and 
\beq
\label{2C}
\chi_{\alpha} = \C(\mu_{\alpha}) + \phi(\nu_{\alpha}), \;\;\;\; \;\;\;\; \alpha =1,...,4,
\eeq
where $\mu_{\alpha}$ and $\nu_{\alpha}$ are phase space functions determining the second class constraints associated to this description. In this representation, the constraints are completely split into first and second class constraints and the constraint matrix \eqref{constmat} reduces (weakly) to 
\beq
\left( \begin{array}{cc} 
0 & 0  \\
0 & \{ \chi_{\alpha} , \chi_{\beta} \}
      \end{array} \right). \nn
\eeq
There is a great deal of freedom in the choice of the $(\mu_{\alpha},\nu_{\alpha})$ coefficients which reflects the non-uniqueness of the above separation of the constraints into first class and second class. However, these phase space functions should be such that
the family $\{(0,k), (X',\partial g g^{-1}), (\pi, [\partial g g^{-1},k]), (\mu_{\alpha},\nu_{\alpha})_{\alpha = 1,..., 4}\}$ of seven-dimensional vectors is linearly independent. 
This implies that the transformation mapping the original set of constraints into the one used here is invertible, and therefore that this new description of the constraint surface is well defined. 
Note that this property ensures that the matrix $\chi_{\alpha \beta} = \{ \chi_{\alpha} , \chi_{\beta} \}$ is invertible. Indeed, we know from the Lagrangian analysis that there are three gauge symmetries in the system and we have found three first class constraints. Therefore, under the assumption that no accidental symmetries are present, there are no further first class constraints in the system. This implies that the constraints $\chi_{\alpha}$ are pure second class and the rank of the matrix $\chi_{\alpha \beta}$ is four, if the coefficients $(\mu_{\alpha},\nu_{\alpha})$  are chosen such that the family of vectors given above is linearly independent.

\medskip

In this general form, the inversion of the above matrix could turn out to be tedious, and consequently, the Dirac bracket could turn out to be un-tracktable. To circumvent this issue, we exploit the ambiguity in the determination of the second class constraints to choose a more specific form for the coefficients appearing in \eqref{2C}. Let $(\mu_{\alpha}, \nu_{\alpha})$ be such that 
the four second class constraints are given by
\beq\label{2C'}
\C(\mu_1),\;\;\; \C(\mu_2), \;\;\; \phi(g' g^{-1}), \;\;\; \phi([g' g^{-1},k]),
\eeq
where $\mu_1$ and $\mu_2$ are two distinct vectors in the four-dimensional target space $\mathbb{M}$ chosen to be orthogonal to both $X'$ and $\pi$.
It is clear that this ansatz does not spoil the linear independency of the family of vectors given above. We chose to order the second class constraints as follows
\beq
\chi_{\alpha} = (\varphi(\mu_i), \phi(\nu_i)) := (\C_i , \phi_i), \;\;\;\; \;\;\;\; i =1,2,
\eeq
where $\nu_1 = g' g^{-1}$ and $\nu_2 = [g' g^{-1} , k]$.
Correspondingly, the four-by-four antisymmetric matrix $\chi_{\alpha \beta}$ now
admits the following block structure
\beq
\chi_{\alpha \beta} \approx  \left( \begin{array}{cc} 
A & B  \\
-B^t & 0 \end{array} \right),
\eeq
where
$$
A_{ij} = - \mu_{[i}^I \mu^J_{j]} M'_{IJ}, \;\;\;\; \mbox{and} \;\;\;\; B_{ij} = \mu_{i}^I \nu_{j}^a (g^{-1} [T_a , k] g)_{IJ} X'^J := \mu_{i}^I \omega_{j I}.
$$
The matrix $\chi_{\alpha \beta}$ is invertible if and only if $\det B \neq 0$. The determinant of $B$ vanishes if $\mu_i^I $ belongs to the plane generated by $(X^I,\pi^I)$, or if $\nu_i^a = k_a$, for all $i=1,2$. Both possibilities are excluded from our ansatz and, as a consequence, $B$ is invertible. 

We can now invert the second class constraint matrix and we obtain the associated Dirac matrix\beq
D_{ij}:=  (\chi^{-1})_{ij} \approx  \left( \begin{array}{cc} 
0 & -(B^t)^{-1}  \\
B^{-1} & B^{-1} A  (B^t)^{-1}  \end{array} \right),
\eeq
where $B^{-1}$ is given by the following expression
$$
B^{-1} = -\frac{1}{\det B} \epsilon \, B^t \, \epsilon,
$$
with $\epsilon$ the two-dimensional totally antisymmetric tensor normalised as in the first part of te paper. Finally, we introduce the Dirac bracket for any couple $(f,g)$ of $C^1$ functions on phase space. It is defined, as usual, by the expression
\beq
\{f,g\}_D = \{f,g\} - \Delta(f,g),
\eeq
where the extension $ \Delta(f,g) = \{f,\chi_{i}\} D^{ij} \{\chi_{j},g\}$ is given explicitly by
\beqa
\Delta(f,g) &=& - \{f,\C_{i}\} ((B^{t})^{-1})^{ij} \{\phi_{j},g\} + \{f,\phi_{i}\} (B^{-1})^{ij} \{\C_{j},g\} \nn \\ && + \{f,\phi_{i}\} (B^{-1} A  (B^t)^{-1})^{ij} \{\phi_{j},g\}.
\eeqa
Although complicated, this bracket has the property that $\{f,g\}_D = \{f,g\}$ for all functions depending only on $X$ and $\pi$ because such functions commute (at least weakly) with the $\phi_i$ constraints. As we are about to see, the partially reduced phase space of the theory can be parametrised by $X$ and $\pi$ only which implies that the Dirac bracket trivialises.

\medskip

We conclude this section by the description of the physical  phase space. 
As we have explicitly computed the Dirac bracket, we can set the second class constraints to zero. 
A first consequence is that the variable $M$ can be solved for in terms of $X$ and $\pi$. Indeed,  the two second class constraints $\C_{i}$ are obtained by projecting the constraint equations $\C_I$ onto two directions which do not belong to the plane formed by $X^I$ and $\pi^I$, which contains the directions where $M$ disappears from the equations. There are therefore two equations for $M$ which needs precisely two numbers to be uniquely determined (it is a self-dual element, invariant under the $U(1)$ group stabilising $k$). Therefore, the second class constraints $\C_{i}$ determine $M$ as a function of $X$ and $\pi$ and of the phase space functions $\mu_i$, at least formally. If we assume that $\mu_i$ is a function only of $X$ and $\pi$, which does not contradict any of our assumptions, $M$ is completely determined as a function of $X$ and $\pi$.

The variable $M$ also parametrises the reduced phase space with respect to the first class constraint $H$ (this partial reduction is consistent because of the structure of the constraint algebra). This is because the set of $H$-orbits is simply given by the conjugacy class of the element $k$ which is parametrised by the elements $M=g^{-1}kg$. As a consequence, on the partially reduced phase space we can set the constraint $H$ explicitly to zero and replace the variable $g$ by $M$. In turn, we have seen that $M$ can be expressed as a function of $X$ and $\pi$ and can therefore be eliminated from the formalism. 

We are therefore left with the eight dimensional phase space parametrised by $X^I$ and $\pi_I$ 
satisfying the standard Poisson algebra
$$
\{ \pi_I(x) , X^J(y) \}_D = \delta_I^J \, \delta(x,y)
$$
since the Dirac structure computed above reduces to the Poisson bracket when evaluated on functions of $X$ and $\pi$ only.
The physical phase space is obtained by the quotient of the action of the two first class constraints 
\begin{eqnarray}
H_1 = \pi_I X'^I  \;\text{and} \;\;\;\; H_0 = \left(\pi_I \pi^I - \frac{\sigma^2}{(2 \pi \alpha')^2} X'_I X'^I \right) \,.
\end{eqnarray}
The parts $\phi(g' g^{-1})$ and  $\phi([ g' g^{-1}, k])$ appearing in \eqref{H0H1} vanish identically because we have solved the two corresponding second class constraints $\phi_i$. The constraint algebra satisfied by $H_1$ and $H_0$ reads
\beqa
\{H_1(u),H_1(u')\}_D \, &=& \, 2 H_1(udu'-u'du)\label{H1H1} \nn \\
\{H_1(u),H_0(v)\}_D \, &=& \, 2 H_0(udv - vdu)\label{H1H0} \nn \\
\{H_0(v),H_0(v')\}_D \, &=& \,-\frac{2 \sigma^2}{(2\pi\alpha')^2} H_1(vdv'-v'dv) \label{H0H0}, \nn
\eeqa 
where, again, the Dirac bracket reduces to the original Poisson bracket because of the first class nature of the constraints. 
We have therefore recovered the physical four-dimensional phase space of the standard Nambu-Goto string and the self-dual string is equivalent to the Nambu-Goto string.

\subsection{Hamiltonian aspects of the general algebraic string}

The Lagrangian analysis displayed in the first part of the paper showed that the general algebraic string contains distinct sectors of solutions, each corresponding to Nambu-Goto strings with different coupling constants. 

\medskip

In this section, we discuss Hamiltonian aspects of the general algebraic string. A canonical analysis of the non-self-dual case was performed in \cite{Us} where it is shown that the algebraic string contains three degrees of freedom per worldsheet point (in configuration space), that is, one extra degree of freedom than the standard bosonic string. The reason for this discrepancy remained unexplained. Here, we provide some further explanation for the appearance of this extra degree of freedom. 

\medskip

The set of constraints of the general theory admits the same generic structure as its self-dual counterpart (\ref{constraints}): it contains the four constraints $\C_I$ and six constraints $\phi_{\alpha \beta}$ instead of three in the self-dual formulation because the full spin algebra is six-dimensional. However, there are some subtle differences.
In the self-dual formulation, two out of the four constraints $\C_I$ turn out to be  totally independent
of $M$ ($\C(\pi)$ and $\C(X')$). The two remaining constraints allow to express completely the two degrees of freedom of $M$ in terms of the coordinates $\pi^I$ and $X'^I$. As a result $M$ can be eliminated from the Hamiltonian theory and this leads directly to the equivalence to the Nambu-Goto string. In the general case, the situation is different;  the four constraints $\C_I$ are not sufficient to express the four degrees of freedom of $M$ in terms of the standard worldsheet variables. There is at least one constraint among the $\C_I$ which is totally independent of $M$;  the spatial diffeomorphism constraint $\C(X')$. The three remaining constraints mix the four degrees of freedom of $M$ with the variables $X'^I$ and $\pi^I$ and there is no way to express the former variables in terms of the latter. 

\medskip

Among these remaining constraints, one (second class constraint) is
particularly interesting:
\beq\label{modifiedH0}
\C(\pi) \; = \; \pi_I \pi^I \, + \, X'^I M_{IJ}^2 X'^J \; \simeq \; 0 \,.
\eeq
For this constraint to reduce to the standard Hamiltonian constraint of the bosonic string, one would need to be able to find a diagonal form of $M^2$ with eigenvalues related to the string tension. More precisely, this diagonal form would need to be a scalar matrix, that is, possess only one unique eigenvalue, as a mean to suppress the dependence on the matrix of eigenvectors (it is not possible to absorb this matrix in a canonical transformation without affecting the diffeomorphism constraint $\C(X')$). To study the diagonalisation of the $M^2$ matrix
we firstly remark the following identities
$$
M^2 \, = \, (k_+^2+k_-^2) \, +  \, 2M_+M_- \;\;\;\; \text{and} \;\;\;\; 
M^4 \, = \, -(k_+^2-k_-^2)^2 \, + \, 2(k_+^2+k_-^2)M^2
$$
with $M=M_++M_-$ the self-dual/anti self-dual decomposition of $M$. We have used that $M_\pm^2=k_\pm^2 = -(1/2) \langle k_{\pm}^2 \rangle \mathbb I$ is proportional to the identity by virtue of \eqref{identity} and the fact that the self-dual and anti-self-dual sectors commute. These identities lead
directly to the following polynomial matrix relation
$$
M^4 +(\langle k_+^2 \rangle + \langle k_-^2 \rangle )M^2 + \frac{1}{4}(\langle k_+^2 \rangle - \langle k_-^2 \rangle)^2 \mathbb I \; = \; 0
$$
which gives by Caley-Hamilton theorem the caracteristic polynomial of  $M$. Namely the eigenvalues $\lambda$
of $M$ are the solutions of the equation
\beq
\lambda^4 + \tau \lambda^2 + \frac{s^2}{4 \sigma^2}\; = \; 0,
\eeq
where we used the relations
$\tau = \langle k_+^2 \rangle + \langle k_-^2 \rangle$ and $s = \sigma (\langle k_+^2 \rangle - \langle k_-^2 \rangle)$. We therefore recover equation (\ref{equation for C}) satisfied by the coupling constant $C(k)^2$ by setting $C(k)^2 = - \sigma^2 \lambda^2$. Accordingly, we find that the matrix $M^2$ has a single eigenvalue if and only if it is self-dual or anti-self-dual, i.e., when $\tau \pm \sigma s = 0$.

\medskip

This result implies that if we are not in the self-dual setting, $M^2$ cannot be eliminated from the constraint \eqref{modifiedH0} and it is not possible to recover the Hamiltonian constraint of the standard bosonic string from \eqref{modifiedH0}. In fact, the standard Hamiltonian constraint can not be obtained as any linear combination of the $\C_I$, unless $M$ is self-dual or anti-self-dual in which case it is given by $\C(\pi)$. It is interesting to remark that the equation satisfied by the eigenvalues of $M^2$ is exactly the equation satisfied by the coupling constant $C(k)$. This analogy leads to the conclusion that the $M^2$ matrix mixes the different sectors of solutions $(\epsilon, \epsilon')$ found in the Lagrangian analysis; each eigenvalue of $M^2$ corresponds to a coupling constant, and thus to a string tension. It is hence tempting to interpret the extra degree of freedom found in \cite{Us} as a dynamical variable interpolating between the possible string tension
 s.

\section{Conclusion and Perspectives}

\subsubsection*{The classical algebraic string}

In this article, we have clarified the link between the algebraic and NG strings at the classical level. We have pursued the work initiated in \cite{Us} by extending the results to both Euclidean and Lorentzian signatures and solving an apparent paradox regarding the relation to the NG string. 

\medskip

We have carried out a careful Lagrangian analysis of the general algebraic string action and showed that the space of solutions to the equations of motion for the first order fields is partitioned into four sectors related by a $\mathbb{Z}_2 \times \mathbb{Z}_2$ symmetry. Evaluating the action on the solutions leads to the conclusion that each sector is equivalent to a NG string with a sector-dependent tension. We have then showed that the four sectors collapse to two sign-related sectors if and only if the first order fields are self-dual or anti-self-dual. Therefore, unless some extra constraints are added by hand to the algebraic string action, only the self-dual string is classically equivalent to the NG string. This explains the first part of the paradox; the anomaly in the number of degrees of freedom described by the general algebraic string found in \cite{Us} is due to the fact that the theory is in a sense larger that the NG string theory, unless one either finds a way of implementing the constraints added by hand to the action in the Hamiltonian framework or restricts to the self-dual framework. 

\medskip

We have then performed a Hamiltonian analysis of the self-dual string and implemented a different strategy to extract the first class constraints 
of the system than in \cite{Us}. This allows to obtain an equivalent description of the constraint surface in which the first class and second class constraints are 
clearly separated. In this new description, the second class constraints are chosen such that when solved strongly, which is equivalent to computing the associated 
Dirac bracket, the resulting phase space is parametrised only by the coordinates and momenta of the NG string. This allows to obtain a description of the physical 
phase space of the self-dual string  which is exactly equivalent to the physical phase space of the NG string. As a consequence, we have proved the equivalence of 
the self-dual and NG strings both at the Lagrangian and Hamiltonian levels. We conclude this paper by a presentation of a new idea to quantise the bosonic string using loop quantum gravity (LQG) techniques. The details will appear elsewhere.

\subsection*{Quantisation: loops vs. strings}

It is essential that any approach to quantise gravity in four-dimensions can be applied to simpler systems that 
we know how to quantise using standard methods. It is not necessary that this approach leads to the same results as
the one obtained with standard techniques but it should at least allow to answer the same questions and to give a clear
description of the quantum theory: physical Hilbert space, scalar product, observables etc. ... .
Examples of such systems are for instance provided by three-dimensional gravity \cite{NouiPerez}, parametrised field theory \cite{Kuchar}, \cite{PFT} or 
the bosonic string.

\medskip

We are mainly interested in the bosonic string, or more precisely the study initiated by Thiemann \cite{Thiemann} consisting in applying LQG techniques to 
quantise the closed bosonic string regarded as a two-dimensional diffeomorphism invariant system.
In particular, we would like to stress the possibility of constructing a vacuum state which is  invariant under diffeomorphisms without being necessarily weakly
discontinuous. Such a state is particularly interesting because it satisfies good properties
for both standard QFT quantisations (the continuity) and LQG (diffeomorphisms invariance). As a consequence, 
the quantisation of the string defined with such a  state is neither equivalent to the Fock space quantisation nor to the quantisation \`a la Thiemann. 
We restrict our discussion here to
the definition the new vacuum state and state some of its properties. We will mainly use the notations introduced in the paper \cite{Policastro}.

\medskip

It is well known that quantisation of the (Lorentzian) bosonic string leads to considering two commuting copies of the (infinite dimensional) 
Weyl algebras $\cal A$. As a vector space, $\cal A$ is  a priori isomorphic to the space  of real valued functions on the string $S^1$ and we will
restrict it to be isomorphic to $C^\infty(S^1, \mathbb R)$: to any function $f \in C^\infty(S^1, \mathbb R)$, one associates an element
$W(f) \in \cal A$. The algebraic structure of $\cal A$ is defined by the relations
\beq
W(f) \, W(g) \; = \; e^{\frac{i}{2}\sigma(f,g)} W(f+g),
\eeq
where the phase is given by $\sigma(f,g)=\int_{S^1} fdg$.  $\cal A$ in naturally endowed with a $C^*$-algebra structure defined
in particular by the conjugation $W(f)^*=W(-f)$.  Furthermore, the group of string reparametrisations defines an automorphism group of
the algebra $\cal A$ in a trivial way: given a reparametrisation $\alpha:S^1\rightarrow S^1$ of the string, one defines
an automorphism, abusively denoted $\alpha$ as well, on $\cal A$ by the relation $\alpha(W(f))=W(f\circ \alpha)$. This map is clearly
an automorphism because $\sigma(f,g)=\sigma (f\circ\alpha,g\circ\alpha)$.
This automorphism group is 
crucial in string theory because it is a signature of the symmetries under diffeomorphims in the language of $C^*$-algebras. 

In the GNS construction, finding the representations of $\cal A$ reduces basicly to findind positive states in $\cal A$:
a state is essentialy a positive linear form $\omega:{\cal A}\rightarrow \mathbb C$ on $\cal A$. 
Two different choices have been proposed for the string so far.
\begin{enumerate}
 \item The Fock space state $\omega_F(W(f))=\langle 0 \vert \pi(W(f)) 0\rangle $ where $\pi:{\cal A} \rightarrow {\cal F}$ is a Fock 
representation of the string and $\vert 0 \rangle$ the vacuum in this representation. 
It is weakly continuous (in the argument $f$) but not invariant under reparametrisations.
 \item The Thiemann state $\omega_T(W(f))=0$ if $f\neq 0$ and  $\omega_T(W(0))=1$. It is trivially invariant under reparametrisations
but not weakly continuous. It presents many similarities with the Ashtekar-Lewandowski state in LQG.
\end{enumerate}
What about finding a state which is weakly continuous and invariant under diffeomorphisms? Such a state $\omega(W(f))$
would depend only on the diffeomorphism class of the function $f$. Physically, one can see that, if $f$ is at least $C^1$,
then it is always possible to find a diffeomorphim $\alpha$ such that $f\circ \alpha$ is a piecewise affine function: this
piecewise affine function is essentially caracterised by the values $(f_i)_i$, $i\in [1,n]$ of its $n$ extrema. Therefore,
a diffeomorphism invariant state $\omega$ depends a priori  only on the values $(f_i)_i$ of its extrema and we can write
$\omega(W(f))=\omega(f_1,\cdots,f_n)$. The Thiemann state is a singular particular state  of this type. But it is possible to
imagine states which are invariant and continuous. For instance, one can propose the following one:
\beq\label{newstate}
\omega(W(f)) \; = \; \exp (-\parallel f \parallel_\infty^2),
\eeq
where $\parallel f \parallel_\infty=\vert \text{Sup}_{S^1}f\vert=\text{Sup}(f_1,\cdots,f_n)$ is the infinite norm of the function $f$.
This last state appears to be particularly interesting because it is both invariant under diffeomorphisms and continuous. It remains to study its associated representation in more details to understand how much it differs from the representations associated to $\omega_F$ and $\omega_T$.
There is a natural generalisation of the state $\omega$ given by the two-parameters family of states 
$$
\omega_{\lambda,N}(W(f)) \; := \; \exp(-\lambda \parallel f \parallel_N^2),
$$ 
where $\lambda$ is a positive number and $N$ is a positive integer which label the $N$-norm $\parallel f \parallel_N$ on the space
of functions on the string. When $N$ and $\lambda$ are finite, $\omega_{\lambda,N}$ is  invariant.under diffeomorphims. However,
the two diffeomorphisms invariant states we have considered so far are obtained from $\omega_{\lambda,N}$ sending one of two parameters
to infinity:
$$
\omega_T=\lim_{\lambda \rightarrow \infty} \omega_{\lambda,N} \;\;\;\;\text{and}\;\;\;\;\;
\omega=\lim_{N \rightarrow \infty} \omega_{\lambda,N} \,.
$$
In the definition of $\omega$, we fixed $\lambda=1$ which is just a matter of convenience.
In that sense, $\omega$ appears as a continuous regularisation of the Thiemann state. 

\medskip

It seems to us very interesting to underline the possibility to construct weakly continuous diffeomorphism invariant representation of the bosonic string.
Of course, so far we have just proposed a state and many properties remain to be studied in great details, the first one being the positivity. First observations
seem to show that the new state is indeed positive. But, even if this is indeed the case, we need to study the representation in details and in particular to find
the spectrum of such a quantum string. We hope to answer these questions in a future article.
 

\appendix 

\section{The Lie algebra $\spin(\eta)$}

In this Appendix, we establish the conventions and notations that are used throughout the paper.

A basis of $\spin(\eta) \cong \so(\eta)$ is provided by the six generators $(T_{\alpha \beta})_{\alpha<\beta = 0,...,3}$. The 
Lie algebra structure is coded in the brackets
\beq
[T_{\alpha \beta}, T_{\gamma \delta} ] = - \eta_{\alpha \gamma} \, T_{\beta  \delta} + \eta_{\alpha \delta} \, T_{\beta \gamma} + \eta_{\beta \gamma} \, T_{\alpha \delta} - \eta_{\beta \delta} \, T_{\alpha \gamma}. 
\eeq
We will make use of the linear Hodge map 
\beq
\label{hodge}
\star : \spin(\eta) \rightarrow \spin(\eta) \, ; \;\;\;\; T_{\alpha \beta} \mapsto (\star T)_{\alpha \beta} = \frac{1}{2} \epsilon_{\alpha \beta}^{\;\gamma \delta} T_{\gamma \delta}
\eeq
where $\epsilon_{\alpha \beta \gamma \delta}$ is the four-dimensional Levi-Cevita tensor normalised by $\epsilon_{0123} = 1$. 
Indices are lowered and raised with $\eta$.
This map is an anti-involution (resp. involution) in Lorentzian (resp. Euclidean) signatures, that is, satisfies $\star^2 = \sigma^2 1 \!\!1$.

It is often convenient to decompose any infinitesimal Lorentz transformation (resp. rotation) into a purely spatial rotation and a hyperbolic rotation, or boost. This is achieved by introducing the rotation and boost generators respectively given by
\beq
J_a = - \frac{1}{2} \, \epsilon_{a}^{\;\, bc} T_{bc}, \;\;\;\; \mbox{and} \;\;\;\; K_a = - T_{a0}, \;\;\;\;\;\;\;\; a,b,c = 1,2,3,
\eeq
where $\epsilon_{abc}$ is the three-dimensional Levi-Cevita tensor such that $\epsilon_{123}=1$.

Using the Lie algebra structure of $\spin(\eta)$ displayed above, it is immediate to check the following commutation relations between the rotation and boost generators
\beq
[J_a, J_b] = \epsilon_{ab}{}^{ c} J_c, \;\;\;\;\;\; [J_a, K_b] = \epsilon_{ab}{}^{c} K_c, \;\;\;\;\;\; [K_a, K_b] =  \sigma^2 \epsilon_{ab}{}^{c} J_c.
\eeq
Furthermore, the action of the Hodge dual on these generators reads $\star K_a = - J_a$.

Via the eigenspace decomposition of the Hodge operator \eqref{hodge}, 
the Lie algebra $\spin(\eta)$ can be decomposed into self-dual and anti-self-dual algebras. 
For Euclidean signatures, the factorisation is merely a change of basis, while for Lorentzian signatures the Hodge squares to minus one and a complexification procedure is required. Let $\g(\sigma)$ be the Lie algebra defined by $\g(\sigma) = \spin(\eta)$ in Euclidean signatures $(\sigma = 1)$ and $\g(\sigma) = \spin(\eta)^{\cC}$ in Lorentzian signature $(\sigma = i)$.

The semi-simple Lie algebra $\g(\sigma)$ splits into two commuting Lie algebras
\beq
\label{factorisation}
\g(\sigma) \cong \h_+(\sigma) \oplus \h_-(\sigma),
\eeq
with $\h_{\pm}(\sigma)$ the simple Lie algebras defined by $\h_{\pm}(1) = \su(2)$ and $\h_{\pm}(i) = \sl(2,\cC)$.
The associated Lie group $SU(2)$ and $SL(2,\mathbb C)$ are denoted $H_\pm(\sigma)$ with $\sigma \in \{1,i\}$.

By virtue of this factorisation, any element $X$ in $\g(\sigma)$ decomposes as
$$
X=X_+ \oplus X_-,  \;\;\;\; \mbox{with} \;\;\;\; X_{\pm} = \frac{1}{2} (\mp \sigma X - \star X).
$$
The self-dual and anti-self-dual components satisfy $\star X_{\pm} = \pm \sigma X_{\pm}$. Using this decomposition,
we can construct the generators of the self-dual and anti self-dual components of $\h_{\pm}(\sigma)$ as follows:
\beq
\label{iso}
T_{\pm a} = \frac{1}{2} \left( \mp \sigma K_a + J_a  \right),
\eeq
It is straightforward to shox that the self-dual and anti-self-dual generators $(T_{\pm a})_a$ satisfy
\beq
[ T_{\pm a} , T_{\pm b} ] = \epsilon_{ab}^{\;\;\,c }T_{\pm c}, \;\;\;\; \mbox{and} \;\;\;\; [T_{+ a}, T_{- b}] = 0,
\eeq
which shows that the split \eqref{factorisation} occurs at the level of Lie algebras.

The spin group $\Spin(\eta)$ acts on $V_{\eta} = \R^{3,1}$ (resp. $V_{\eta} = \R^4$) via the homomorphism to $\SO(\eta)$. Let $(\pi, V_{\eta})$ denote the corresponding (vector)  representation and $(e_I)_{I=0,...,3}$ be a choice of basis of $V_{\eta}$. The induced representation $\pi_*:\spin(\eta) \rightarrow \End (V_{\eta})$ of the Lie algebra is defined by the following evaluations 
\beq
\label{vector}
\pi_*(T_{ab})^I_{\;\;J}:= (T_{ab})^I_{\;\;J} = \delta_a^I \eta_{bJ} - \eta_{aJ} \delta_b^I,
\eeq
where $\delta$ is the Kronecker symbol. It is then immediate to obtain the image of the self-dual and anti-self-dual generators in the vector representation
$$
\pi_*(T_{\pm a})^I_{\;\;J} = \frac{1}{2} \left( \mp \sigma (\delta^{I}_0 \eta_{aJ} - \eta_{0J} \delta_a^{I}) -\epsilon_{0a\;J}^{\;\;\;I} \right).
$$
We will make use of the following relation
\beq
\label{identity}
\pi_*(T_{\pm a}) \circ \pi_*(T_{\pm b}) = \frac{1}{2} \epsilon_{ab}^{\;\;\;c} \, \pi_*(T_{\pm c}) - \frac{1}{4} \delta_{ab} 1\!\!1.
\eeq
Note that the left hand side is not necessarily an antisymmetric matrix since the term is quadratic and thus in $U(\spin(\eta))$ not in the Lie algebra $\spin(\eta)$.

Finally, we introduce the two symmetric, $\Ad$-invariant, non-degenerate bilinear forms on $\spin(\eta)$ which are defined by
\beq
\langle T_{\alpha \beta}, T_{\gamma \delta} \rangle = \eta_{\alpha \gamma} \eta_{\beta \delta} - \eta_{\alpha \delta} \eta_{\beta \gamma}, \;\;\;\;
\text{and} \;\;\;  (T_{\alpha \beta}, T_{\gamma \delta} ) =  \epsilon_{\alpha \beta \gamma \delta}.
\eeq
Using the expression for the matrix elements of the generators, it is straightforward to relate the two bilinear forms to the trace in the vector representation `$\mathrm{Tr}$' and obtain that $\la , \ra = -\frac{1}{2} \mathrm{Tr}$. Evaluated on the boost and rotation generators, the Killing form yields
$$
\langle J_a, J_b \rangle = \delta_{ab}, \;\; \langle J_a, K_b \rangle = 0, \;\; \langle K_a, K_b \rangle = \sigma^2 \delta_{ab}.
$$
From this, it is immediate to see that the self-dual/anti-self-dual decomposition is orthogonal in the Killing form $\langle , \rangle$ which reduces to (one-half times) the Killing form `$\tr$' on each one of the two copies:
$$
\la T_{\epsilon a}, T_{\epsilon'b} \ra = \frac{1}{2} \delta_{ab} \delta_{\epsilon \, \epsilon'},
$$ 
with $\epsilon, \epsilon' = \pm$. We therefore have the following relation between the various bilinear forms and traces introduced above
$$
\langle X, Y \rangle = - \frac{1}{2} \mathrm{Tr} (XY) = \frac{1}{2} \tr (XY),
$$
for every purely self-dual or anti-self-dual elements $X, Y$ in $\g (\sigma)$.

\section{Variational problem for the Nambu-Goto string}

In this Appendix, we recall how one obtains the wave equation for the worldsheet coordinates 
from the Nambu-Goto action. 

As the action depends only on the derivatives $\partial_x X$ and $\partial_t X$ of the
field, the Euler-Lagrange equations reduce to:
\beq
\partial_t\left( \frac{\partial {\cal L}}{\partial \partial_t X}\right) \; + \; \partial_x\left( \frac{\partial {\cal L}}{\partial \partial_x X}\right) \; = \; 0
\eeq
where $\cal L$ is the Lagrangian density and each fonctional derivatives present in the equations of motion take the form
\beq
\frac{\partial {\cal L}}{\partial \partial_t X} \; = \; \gamma_{xx} \partial_t X \, - \, \gamma_{xt} \, \partial_x X \;\;\;\;\;\;
\frac{\partial {\cal L}}{\partial \partial_x X} \; = \; \gamma_{tt} \partial_x X \, - \, \gamma_{xt} \, \partial_x X\,.
\eeq
The coefficients $\gamma_{ij}$ are the components of the normalised induced metric $\gamma$ on the worldsheet
according to:
\begin{eqnarray}
\gamma \; := \; \frac{1}{\sqrt{\vert \det X^* \eta\vert}} X^* \eta \; = \; \left( \begin{array}{cc} \gamma_{tt} & \gamma_{tx}\\ 
\gamma_{tx} & \gamma_{tt} \end{array} \right)\,.
\end{eqnarray}
Therefore, they satisfisfy the relation $\gamma_{tt}\gamma_{xx}-\gamma_{tx}^2=\sigma^2$. As a consequence, the equations of motion read
\beq
\partial_t(\gamma_{xx}\partial_t - \gamma_{tx} \partial_x)X \, + \, \partial_x(\gamma_{tt} \partial_x-\gamma_{tx} \partial_t)X \, = \, 0 \,.
\eeq
It is a priori not obvious to see from these equations that the field $X$ are harmonic functions but changing the time vector-field
$\partial_t$ by $\tilde{\partial}_t=\gamma_{xx}\partial_t - \gamma_{tx} \partial_x$ leads to the following reformulation of the equations of motion 
\beq
\tilde{\partial}_t^2 X \; + \; \sigma^2 \, \partial_x^2X \; = \; 0 \,.
\eeq
In other words, the choice of a ``good'' time variable allows to obtain the wave equation for the variables $X$
which is what we expected. 

\subsubsection*{Acknowledgements}
KN wants to thank Hanno Salhmann and Guisepe Policastro for discussions concerning the quantisation of the string.
This work was partially supported by the ANR. WF is funded by the Emmy Noether grant ME 3425/1-1 of the German Research Foundation (DFG).

\end{document}